\documentclass[preprint2]{aastex}

\newcommand{\hb}{H$\beta$}
\newcommand{\ha}{H$\alpha$}

\shortauthors{Onken et al.}
\shorttitle{Black Hole Masses in Three Seyfert Galaxies}

\received{}
\accepted{}

\begin{document}

\title{Black Hole Masses in Three Seyfert Galaxies}
\author{C.A.~Onken\altaffilmark{1}, B.M.~Peterson\altaffilmark{1}, 
M.~Dietrich\altaffilmark{2}, A.~Robinson\altaffilmark{3}, \& I.M.~Salamanca\altaffilmark{4}}
\altaffiltext{1}{Department of Astronomy, The Ohio State University, 
140 West 18th Avenue, Columbus, OH 43210; onken@astronomy.ohio-state.edu, 
peterson@astronomy.ohio-state.edu}
\altaffiltext{2}{Department of Astronomy, University of Florida, 211 Bryant Space Science Center,
Gainesville, FL, 32611-2055. Current address: Department of Physics and Astronomy, Georgia
State University, One Park Place South SE, Atlanta, GA, 30303; dietrich@chara.gsu.edu}
\altaffiltext{3}{Division of Physics and Astronomy, Department of Physical Sciences,
University of Hertfordshire, College Lane, Hatfield, Herts AL10 9AB, United Kingdom; 
ar@star.herts.ac.uk}
\altaffiltext{4}{Anton Pannekoek Institute, University of Amsterdam, Kruislaan 403,
Amsterdam 1098 SJ, The Netherlands; isabel@science.uva.nl}

\begin{abstract}

We analyze published reverberation mapping data for three Seyfert galaxies (NGC 3227,
NGC~3516, and NGC~4593) to refine the mass estimate for the supermassive black hole in the center
of each object. Treatment of the data in a manner more consistent with other large
compilations of such masses allows us to more securely compare our results to wider samples of
data, e.g., in the investigation of the M$_{bh}$-$\sigma_{\ast}$ relationship for active and quiescent
galaxies.

\end{abstract}


\clearpage

\section{INTRODUCTION}

Reverberation mapping \citep{bla82,pet93} is an important technique 
for probing the supermassive black holes (SMBHs) 
at the centers of active galactic nuclei (AGNs). Measuring the time lags, $\tau$,
between fluctuations in the highly variable AGN continuum and the variations of 
emission lines arising from
the broad-line region (BLR) provides a responsivity-weighted radius for the
BLR gas distribution around the SMBH.
When emission-line time lags are combined with calculations of the gas velocity
width, $\sigma$, under the assumption of virialized gas motions, it 
becomes possible to estimate the SMBH mass.

Multiple lines of evidence suggest that to assume a virialized BLR is
a reasonable approximation. In AGNs for which several emission lines
have been reverberation-mapped (NGC~5548, 3C~390.3, NGC~7469, NGC~3783), an
inverse relationship is found between $\tau$ and $\sigma$ that is consistent
with expectations from virialization \citep{pet99,pet00,onk02}. In addition,
the close concordance between AGNs and quiescent galaxies in the 
$M_{bh}$-$\sigma_{\ast}$ plane under the virial assumption 
\cite[relating black hole mass to the stellar 
velocity dispersion of the galactic spheroid;][]{fer01} would seem to imply that any
systematic error in reverberation masses is small (a factor of $\sim$3 or less).
\citet{kro01} estimated a similar level of expected systematic scatter assuming
gravitational domination of the BLR.

Most of the statistical studies of reverberation-based
black-hole masses use the homogeneous compilations of
Wandel, Peterson, \& Malkan (1999, hereafter WPM)
and \citet{kas00}.
Missing from these compilations are AGNs observed
by the ``Lovers of Active Galaxies (LAG)'' collaboration that
monitored several AGNs in early 1990
\citep{rob94}. Three of the LAG sources,
NGC~3227 \citep{sal94}, NGC~3516 \citep{wan93}, and
NGC~4593 \citep{die94}, have well-determined emission-line
lags and certainly meet the quality criteria for inclusion
in these studies, and indeed were included in the compilation
of Ho (1999; based on \hb\ only). These galaxies were omitted from the
WPM compilation only because the data had not been analyzed
in the same fashion as the other sources;
specifically,
(a) WPM use the model-independent Monte Carlo
method of \citet{pet98} to assess uncertainties
in emission-line lags, and
(b) WPM use the FWHM of the emission-line in
the root-mean-square (rms) spectrum, rather than the mean spectrum,
to characterize the BLR line-of-sight velocity
width, $\Delta V$, that is used to form the virial
product ($\Delta V$)$^{2} \tau$.
Because we have converted both our time lags and velocity widths to
the AGN rest frame, our results supplement the compilation of \citet{kas00}.
(WPM did not correct for the time dilation of their $\tau$ values.)

In this contribution, we reanalyze the LAG data
on these three sources in the same fashion as \citet{kas00}, with
the intent of enlarging that homogeneous database.
The total number
of AGNs which have been reverberation-mapped (a few dozen) is small enough that
additional objects will assist in on-going statistical investigations of (what are hoped
to be) fundamental physical relationships. 
The main aspect on which we have worked is the careful construction the 
rms spectra over the spans of the respective observing campaigns.
With the rms spectra for these
AGNs and updated time lag values and uncertainties, we
are able to derive masses for the SMBHs in these three galaxies.

In \S\ 2, we briefly describe the original observing campaigns and our data reduction.
We explain the details of our time series and velocity width analyses in \S\ 3. Our 
estimates of the SMBH mass in each galaxy are
presented in \S\ 4. In \S\ 5, we address the M$_{bh}$-$\sigma_{\ast}$ relationship, and
we then summarize our conclusions (\S\ 6).

\section{OBSERVATIONS AND DATA REDUCTION}

The LAG consortium examined the \ha\ and \hb\ wavelength regimes with long slit 
spectroscopy of moderate resolution and with the temporal sampling rates 
listed in Table \ref{tab1}. The line marked ``cont.'' in Table \ref{tab1} for
NGC~3516 indicates the sampling statistics derived for the combination of imaging
and spectroscopic measurements of the $\sim$4000-5000\AA\ continuum. The continuum
statistics for NGC~3227 and NGC~4593 are identical to the \hb\ and \ha\ entries,
respectively. $F_{var}$ indicates the ``excess variance,'' the mean fractional
variation of each dataset \cite[see][]{rod97}. The ratio of the maximum to minimum
flux value is given in Table \ref{tab1} as $R_{max}$.

In general, no alterations were applied to the published
light curve data. However, in order to investigate the velocity width
characteristics of the broad emission lines, we used the wavelength
region around the narrow lines
([\ion{O}{3}] and [\ion{S}{2}], which should only vary on timescales 
much longer than the campaigns) to scale the spectra \citep{van92}.
The absolute scaling was done so as to yield mean narrow-line fluxes equal
to the values derived in the original papers.
The slit width selected was an attempt to balance spectral resolution and total 
detected flux. However, the small slit width 
introduced a non-trivial degree of difficulty in the removal of seeing
effects, as noted below.

\subsection{NGC~3227}

The LAG campaign on NGC~3227 was described by \citet{sal94}. The analysis was
complicated by the fact that the two methods they employed for removing the contribution of the
stellar host gave significantly different answers: they derived a 24\% contribution to the 
continuum by examining the magnesium triplet of a template galaxy of the 
same Hubble type, and a 40\% continuum contribution from comparisons with the magnesium
triplet absorption of the bulge of NGC~3227. We analyzed the data with each of the
potential stellar levels removed and found the differences to be insignificant.

A detailed investigation of various origins of uncertainty in their flux calibration
was conducted by a portion of the LAG group \citep{bar94}. In addition to the effects
of slit miscentering and seeing, they estimate the uncertainties incurred by their
scaling of the spectra by the narrow lines, and conclude an overall flux error of 
5--8\%. We have assumed the more conservative 8\% errors for all of the flux data.
Our analysis made use of the full LAG dataset, consisting of 23 \ha\ observations
and 14 \hb\ observations. The light curve cross-correlations were made with respect to the
continuum at 5020\AA.

The systemic redshift of NGC~3227 was taken to be $z$ = 0.00380 $\pm$ 0.00002 \citep{kee96}.

\subsection{NGC~3516}

Spectra of NGC~3516 were taken at 22 separate epochs for \ha\ and 21 epochs for \hb. The
continuum light curve for this object was constructed from a combination of B-band imaging
and spectral fits to the power-law continuum at 4905\AA.
The results of the LAG study have been published by \citet{wan93} and \citet{wan94}.
The spectra we obtained for this object had already been corrected for seeing,
the stellar flux from the host galaxy, and contributions from narrow lines
\cite[see][]{wan92}. In addition, the [\ion{S}{2}] and [\ion{O}{3}] narrow lines 
had been subtracted from the \ha\ and \hb\ spectra, respectively, preventing any 
rescaling of the flux levels.
Several observations were removed from the calculation of the rms spectrum
for a variety of reasons (extremely poor seeing, object miscentering, and other
anomalies).
Among the \ha\ dataset, we have excised observations from the following dates:
31 January, 16 February, 15 April, and 13 May. From the \hb\ analysis 
we have removed 24 March, 15 April, and 13 May.

Conversion of time lags and line-of-sight velocity widths to the rest frame
was made with $z$ = 0.00884 $\pm$ 0.00002 \citep{kee96}.

\subsection{NGC~4593}

Spectra of NGC~4593 were obtained at 22 epochs around \ha\ and at 11 epochs around \hb, 
and the continuum used was at 6310\AA.
The \ha\ line was contaminated by strong [\ion{N}{2}] $\lambda\lambda$6548, 6584 emission, 
as well as a narrow \ha\ contribution.
Using the shape of the narrow [\ion{O}{3}]$\lambda$5007 line as a template for all of the
narrow lines, and the flux ratios between the [\ion{N}{2}] lines and narrow \ha\ that 
\citet{die94} calculated, we have bracketed the amount of narrow line removal by limiting cases
in which too much or too little narrow line flux was subtracted. Conducting this for
each \ha\ spectrum, we then separately analyzed our oversubtracted and undersubtracted
datasets. The $\Delta V$ difference between the oversubtraction and undersubtraction 
was on the order of 1\%, much smaller than the $\Delta V$ uncertainties as well 
as the range of values
derived by doing no rescaling of the spectra or no removal of the narrow-line contribution.

We converted our results to the galaxy's rest frame using $z$ = 0.00900 $\pm$ 0.00013 \citep{str92}. 

\section{DATA ANALYSIS}

Using the published light curves, we generated the cross-correlation functions (CCFs) between
the continua and the emission lines with the interpolated CCF technique as implemented by 
\citet{whi94}. We calculated 
both the peak of each CCF ($\tau_{peak}$) and its centroid ($\tau_{cent}$;
above a threshold of 80\% of the peak value correlation coefficient) 
with an interpolation unit of 0.1 days.
 
Uncertainties in the time lag values were estimated with the method described by
\citet{pet98}. This model-independent technique addresses two major sources of 
uncertainty in the CCFs. First, the issue of flux errors is handled by adding to
each flux value a randomly generated deviation. This deviation is computed such
that a large number of realizations will produce a Gaussian scatter with a 
standard deviation equal to the quoted uncertainty (this procedure is dubbed 
``flux randomization (FR)''). Second, the potential
errors arising from the limited temporal sampling of the observational campaign are
accounted for by ``random subset selection (RSS).'' The RSS method consists of randomly
selecting $N$ data points from a sample of size $N$, where each selection draws from
the entire sample and repeated selections are ignored. This allows us to ``bootstrap''
our way to reasonable error estimates, while preserving the underlying form of the
time series. The FR/RSS method was implemented 1000 times for each continuum-line combination 
to build up a distribution 
of the peak and centroid time lags, denoted generally as a cross-correlation peak
distribution \cite[CCPD;][]{mao89}. We use the CCPD to define the time lag values within
which 68\% of the Monte Carlo results lie, and these values set our time lag uncertainties.
Table \ref{tab2} and Figure \ref{fig3} show our cross-correlation results.

The procedure we applied to the mean and rms spectra of each AGN was to
make two estimates of the continuum, at the extrema of where
the level could potentially be set. The two resulting measurements of the
velocity full-width at half-maximum, $\Delta V$, were then averaged
and the difference taken to be twice the one-sigma errors.
We give the results of our $\Delta V$ measurements from both the mean and 
rms spectra in Table \ref{tab3}.

\section{MASS DERIVATIONS}

For consistency with previously published compilations, we
calculate the reverberation mass as
\begin{equation}
M_{rev} = \frac{3\ c\ \tau\ (\Delta V)^{2}}{4\ G}
\end{equation}
where $\Delta V$ is taken from the rms spectrum, $c$ is the speed of light, 
and $G$ is the gravitational constant.
The asymmetric errors in the time lags are propagated through to the masses
derived from the individual lines. Then, the masses for each line are 
combined in a weighted average.
We have measured the reverberation masses for our three targets and
present the results in Table \ref{tab4}. Object-specific comments and
comparisons with previous results follow.

\subsection{NGC~3227}

The original LAG estimate of the SMBH mass in NGC~3227 was 
$\sim$2$\times$10$^{8}$M$_{\odot}$ \citep{sal94}. 
Our value of (3.6$\pm$1.4)$\times$10$^{7}$M$_{\odot}$ is consistent with the
mass of 4.9$^{+2.7}_{-5.0}\times$10$^{7}$M$_{\odot}$ tabulated by WPM, which was based
on an independent monitoring campaign \citep{win95}. 
The value derived by \citet{ho99} from \hb\ was 3.8$\times$10$^{7}$M$_{\odot}$, also
consistent with our new SMBH mass in this galaxy.
\citet{sch00} used CO gas kinematics to calculate a lower mass limit of 
$\sim$1.5$\times$10$^{7}$M$_{\odot}$ enclosed within the central 25 pc. Again,
this value is fully consistent with our results.

\subsection{NGC~3516}

\citet{kaz01} have noted the difficulty in reconciling the SMBH mass estimates for
NGC~3516 arising from a variety of optical, UV, and X-ray constraints within certain
models of AGN structure.
Although our final 
estimate of $M_{rev}$ = (1.68$\pm$0.33)$\times$10$^{7}$M$_{\odot}$ lies slightly 
above a recent mass upper limit of (1.12$\pm$0.05)$\times$10$^{7}$M$_{\odot}$
derived from models
of the X-ray warm absorber in this galaxy \citep{mor02}, other models of X-ray 
properties point towards a mass closer to 10$^{8}$M$_{\odot}$ \citep{kaz01}.
Our results for this AGN are consistent with the value of
$\sim$2$\times$10$^{7}$M$_{\odot}$ from the earlier analysis of the LAG data
by \citet{wan94}, and also matches
well the mass of 2.3$\times$10$^{7}$M$_{\odot}$ \citet{ho99} calculated using \hb\ alone. 
Recent attempts at accretion disk modeling from X-ray data have given values of 
3.1$\times$10$^{7}$M$_{\odot}$ \citep{cze01} and 2$\times$10$^{7}$M$_{\odot}$ \citep{chi02},
both in rough agreement with the present work.

\subsection{NGC~4593}

For NGC~4593, we calculate that M$_{rev}$ = (6.6$\pm$5.2)$\times$10$^{6}$M$_{\odot}$.
An earlier reverberation mapping campaign derived a mass of 
2.2$^{+1.4}_{-1.1}\times$10$^{6}$M$_{\odot}$ from Lyman $\alpha$ \citep{san95},
and \citet{kol97} calculated M$_{rev}$ = 7$\times$10$^{6}$M$_{\odot}$ from the LAG \ha\
data (when corrected for our conversion between $\Delta V$ and the velocity dispersion).
\citet{ho99} estimated a slightly larger \hb\ time delay and found a SMBH mass of
8.1$\times$10$^{6}$M$_{\odot}$. With such large uncertainties, these differences
cannot be considered statistically significant.

\section{THE M$_{bh}$-$\sigma_{\ast}$ RELATIONSHIP}

Since the discovery of the exceptionally tight correlation between
a galaxy's SMBH mass and the central value of the velocity dispersion of the galaxy's spheroid
\cite[i.e. M$_{bh}$-$\sigma_{\ast}$;][]{fer00,geb00}, a number of attempts have
been made to reconcile differences in the derived slope of this relationship
\cite[e.g.,][]{tre02} as well as to develop a physical framework that will
naturally produce such a correlation. In addition, \citet{geb00b} and \citet{fer01} 
have examined whether SMBH masses derived by reverberation
mapping are consistent with the M$_{bh}$-$\sigma_{\ast}$ relationship. Because
bulge velocity dispersion measurements exist for all three of our targets, we
are able to enlarge the AGN sample used to investigate M$_{bh}$-$\sigma_{\ast}$.

\citet{nel95} measured bulge velocity dispersions for 85 galaxies, including 
two of our targets. NGC~3227 was found to have $\sigma_{\ast}$ = 128$\pm$13 km s$^{-1}$,
and NGC~4593 has $\sigma_{\ast}$ = 124$\pm$29 km s$^{-1}$. \citet{din95} published a value of 
$\sigma_{\ast}$ = 124$\pm$5 km s$^{-1}$ for NGC~3516. 
These values for $\sigma_{\ast}$, in conjunction with
our derived M$_{rev}$ data, are consistent with the current fits to
M$_{bh}$-$\sigma_{\ast}$ (Figure \ref{fig4}). This excellent agreement between the 
AGN and quiescent galaxy
M$_{bh}$-$\sigma_{\ast}$ relationships is additional evidence that reverberation
masses are not subject to large systematic errors. 

In addition, we note the
relative concentrations of the AGN and quiescent galaxy data in Figure \ref{fig4}
toward the two ends of the relationship.
This is a reflection of two facts: (1) in order to measure $\sigma_{\ast}$ in AGNs, 
the continuum emission must be weak enough
so as not to wash out the stellar absorption lines, implying the accessibility of only the dimmer 
AGNs, and therefore only the smaller SMBH masses; 
(2) the quiescent galaxies need a relatively large SMBH
mass to allow the probing of the gravitational sphere of influence, whether by stellar
dynamics or gas kinematics. The combination of these requirements gives rise to the
observed distribution, in spite of the probable span of both galaxy types over 
the entire range of SMBH masses.

\section{SUMMARY}

As many of the forefront studies of AGNs are examining correlations between
SMBH masses and properties of the AGN host galaxies, a uniform analysis 
methodology is important for minimizing systematic offsets between
datasets.
Time lags, calculated from the published light curves, were used in conjunction with
rms velocity widths for optical hydrogen emission lines to estimate the reverberation
masses for NGC~3227, NGC~3516, and NGC~4593. 
These masses, listed in Table \ref{tab4}, can now be added to the 
compilation of \citet{kas00}, as they were calculated in the same manner. In addition, 
these masses, when combined with published data on the host galaxy spheroid velocity
dispersions, strengthen the claim that reverberation masses are accurate measures
of the true SMBH mass.

\acknowledgments

We acknowledge support for this work through NASA grant NAG5-8397.
C. A. O. thanks The Ohio State University for support through the Distinguished
University Fellowship.
This research has made use of the NASA/IPAC Extragalactic Database (NED) 
which is operated by the Jet Propulsion Laboratory, California
Institute of Technology, under contract with the National Aeronautics 
and Space Administration.

%figures

\clearpage

\begin{figure}
\plotone{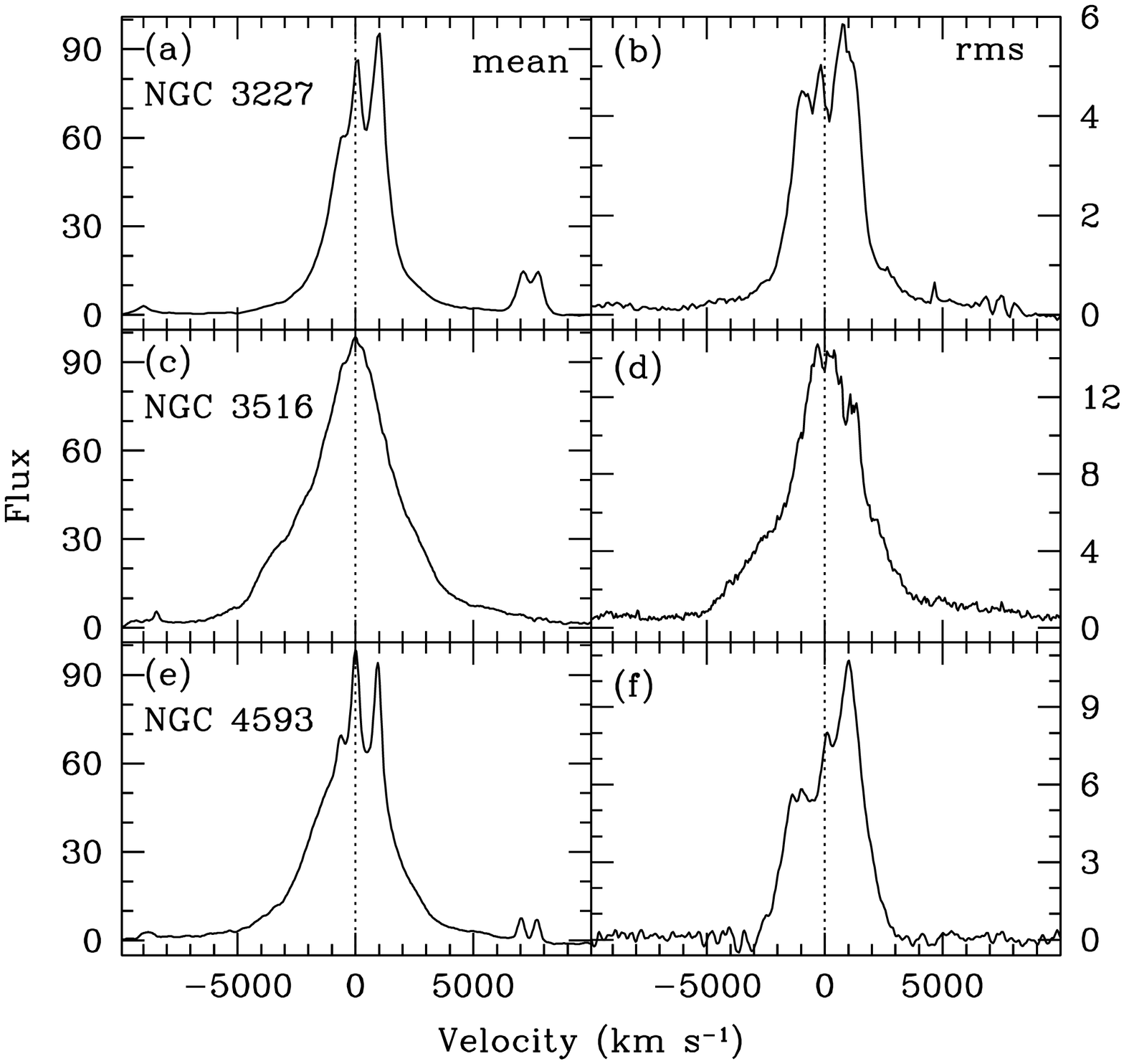}
\caption[f1.eps]{Mean and rms \ha\ spectra for NGC~3227 ({\it (a)} and {\it (b)}),
NGC~3516 ({\it (c)} and {\it (d)}), and 
NGC~4593 ({\it (e)} and {\it (f)}). The y-axis scales of $F_{\lambda}$ are arbitrary and the
continua have been subtracted. \label{fig1}}
\end{figure}

\clearpage

\begin{figure}
%\epsscale{.60}
\plotone{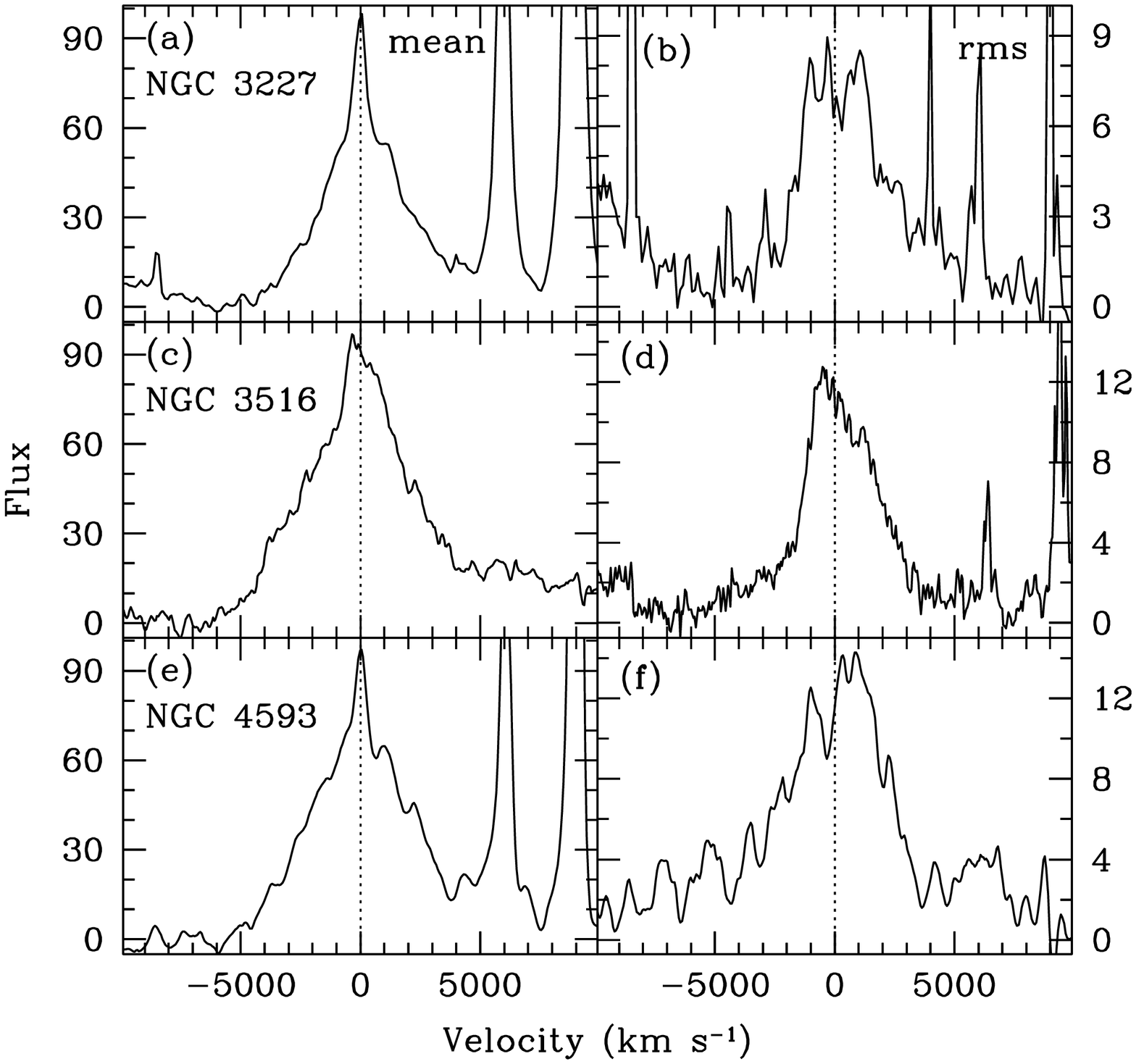}
\caption[f2.eps]{Mean and rms \hb\ spectra for NGC~3227 ({\it (a)} and {\it (b)}),
NGC~3516 ({\it (c)} and {\it (d)}), and 
NGC~4593 ({\it (e)} and {\it (f)}). The y-axis scales of $F_{\lambda}$ are arbitrary and the
continua have been subtracted. \label{fig2}}
\end{figure}

\clearpage

\begin{figure}
\plotone{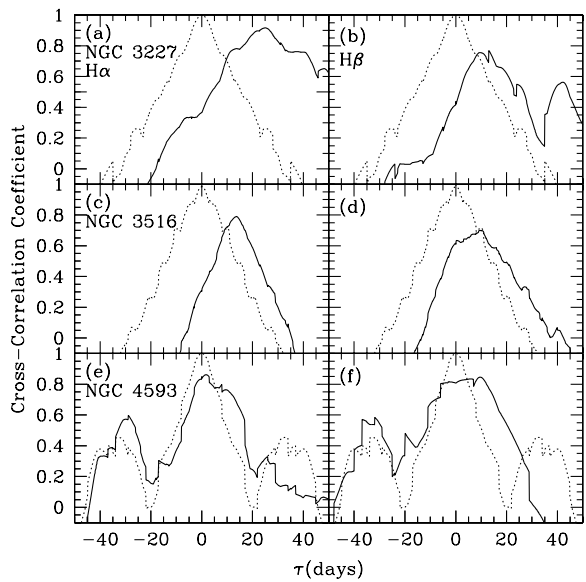}
\caption[f3.eps]{Cross-correlation results for H$\alpha$ (left panels) and H$\beta$ (right panels)
with continuum auto-correlation functions (dotted lines; cross-correlation of the continuum with 
itself) for NGC~3227 ({\it (a)} and {\it (b)}), NGC~3516 ({\it (c)} and {\it (d)}), and NGC~4593 
({\it (e)} and {\it (f)}). \label{fig3}}
\end{figure}

\clearpage

\begin{figure}
\epsscale{.60}
\plotone{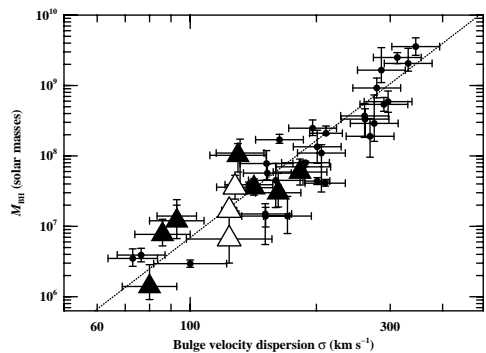}
\caption[f4.eps]{The relationship between black-hole mass and the central
value of host-galaxy bulge velocity dispersion for quiescent and active galaxies.
The filled circles represent quiescent galaxies (data kindly provided by 
L.\ Ferrarese) and the filled triangles are AGNs from \citet{fer01}. The
open triangles are the three AGNs discussed in this paper. The dashed line is the 
best fit to the quiescent-galaxy data, $M_{\rm BH} \propto \sigma^{4.58}$ \citep{fer02}.
\label{fig4}}
\end{figure}

\clearpage

\begin{deluxetable}{lccccccc}
\tabletypesize{\footnotesize}
\tablecaption{Sampling Statistics \label{tab1}}
\tablewidth{0pt}
\tablehead{
\colhead{} & \colhead{} & \colhead{} & \multicolumn{2}{c}{Sampling Interval (days)} &
\colhead{} & \colhead{} & \colhead{} \\
 \cline{4-5} \\
\colhead{Galaxy} & \colhead{Dataset} & \colhead{Number} & \colhead{Average} & \colhead{Median} &
\colhead{F$_{var}$} & \colhead{R$_{max}$}
}
\startdata
NGC~3227 & \ha\ & 23 & 6.9 & 5.1 & 0.017 & 1.367$\pm$0.155 \\
\nodata  & \hb\ & 14 & 10.6 & 10.9 & 0.133 & 1.597$\pm$0.181 \\
NGC~3516 & \ha\ & 18 & 8.9 & 8.8 & 0.129 & 1.541$\pm$0.065 \\
\nodata  & \hb\ & 18 & 9.0 & 8.0 & 0.112 & 1.474$\pm$0.063 \\
\nodata  & cont. & 35 & 4.5 & 3.0 & 0.280 & 4.027$\pm$0.364 \\
NGC~4593 & \ha\ & 22 & 7.4 & 5.0 & 0.136 & 1.653$\pm$0.069 \\
\nodata  & \hb\ & 11 & 15.8 & 11.7 & 0.185 & 1.996$\pm$0.102 \\
\enddata
\end{deluxetable}

\clearpage

\begin{deluxetable}{lllcccc}
\tablecaption{Cross-Correlation Results \label{tab2}}
\tablewidth{0pt}
\tablehead{
\colhead{Galaxy} & \colhead{Emission Line} & \colhead{Continuum} & \colhead{$\tau_{cent}$} &
\colhead{$\tau_{cent}^{rest}$} & \colhead{$\tau_{peak}$} & \colhead{$\tau_{peak}^{rest}$} \\
\colhead{} & \colhead{} & \colhead{(\AA)} & \colhead{(days)} &
\colhead{(days)} & \colhead{(days)} & \colhead{(days)}
}
\startdata
NGC~3227 & \ha\ (24\% stellar) & 5020 & 25.9$_{-10.0}^{+14.9}$ & 25.8$_{-10.0}^{+14.8}$ & 25.0$_{-9.6}^{+18.8}$ & 24.9$_{-9.6}^{+18.7}$ \\
\nodata  & \hb\ (24\% stellar) & \nodata & 12.0$_{-9.1}^{+26.8}$ & 12.0$_{-9.1}^{+26.7}$ & 13.0$_{-13.1}^{+26.6}$ & 13.0$_{-13.1}^{+26.5}$ \\
NGC~3516 & \ha & composite & 13.2$_{-2.6}^{+5.9}$ & 13.1$_{-2.6}^{+5.8}$ & 13.8$_{-3.8}^{+5.3}$ & 13.7$_{-3.8}^{+5.2}$ \\
\nodata  & \hb & \nodata & 7.4$_{-2.6}^{+5.4}$ & 7.3$_{-2.5}^{+5.4}$ & 10.2$_{-6.7}^{+0.9}$ & 10.1$_{-6.6}^{+0.9}$ \\
NGC~4593 & \ha & 6310 & 4.6$_{-5.0}^{+2.5}$ & 4.6$_{-5.0}^{+2.5}$ & 1.7$_{-1.7}^{+7.8}$ & 1.7$_{-1.7}^{+7.7}$ \\
\nodata  & \hb & \nodata & 3.1$_{-5.1}^{+7.6}$ & 3.1$_{-5.1}^{+7.5}$ & 9.6$_{-12.6}^{+1.2}$ & 9.5$_{-12.5}^{+1.2}$ \\
\enddata
\end{deluxetable}

\clearpage

\begin{deluxetable}{lcccccc}
\tablecaption{Velocity Width Results \label{tab3}}
\tablewidth{0pt}
\tablehead{
\colhead{} & \colhead{} & \multicolumn{2}{c}{H$\alpha$} & \colhead{} & \multicolumn{2}{c}{H$\beta$}  \\
\cline{3-4} \cline{6-7} \\
\colhead{Galaxy} & \colhead{Dataset} & \colhead{$\Delta V$} & \colhead{$\Delta V^{rest}$} & \colhead{} &
\colhead{$\Delta V$} & \colhead{$\Delta V^{rest}$} \\
\colhead{} & \colhead{} & \colhead{(10$^{3}$ km s$^{-1}$)} & \colhead{(10$^{3}$ km s$^{-1}$)} & \colhead{} &
\colhead{(10$^{3}$ km s$^{-1}$)} & \colhead{(10$^{3}$ km s$^{-1}$)}
}
\startdata
NGC~3227 & rms     & 3.10$\pm$0.15 & 3.09$\pm$0.15 & & 4.36$\pm$1.32 & 4.34$\pm$1.31 \\
NGC~3516 & \nodata & 3.11$\pm$0.04 & 3.08$\pm$0.04 & & 3.14$\pm$0.15 & 3.11$\pm$0.15 \\
NGC~4593 & \nodata & 3.10$\pm$1.09 & 3.07$\pm$1.08 & & 4.42$\pm$0.95 & 4.38$\pm$0.94 \\
NGC~3227 & mean    & 2.62$\pm$0.22 & 2.61$\pm$0.22 & & 1.90$\pm$0.57 & 1.89$\pm$0.57 \\
NGC~3516 & \nodata & 3.51$\pm$0.10 & 3.48$\pm$0.10 & & 4.09$\pm$0.77 & 4.05$\pm$0.76 \\
NGC~4593 & \nodata & 3.27$\pm$0.66 & 3.24$\pm$0.65 & & 3.31$\pm$0.93 & 3.28$\pm$0.92 \\
\enddata
\end{deluxetable}

\clearpage

\begin{deluxetable}{llll}
\tablecaption{Reverberation Masses \label{tab4}}
\tablewidth{0pt}
\tablehead{
\colhead{Galaxy} & \colhead{H$\alpha$-based} & \colhead{H$\beta$-based} & \colhead{Combined} \\
\colhead{} & \colhead{(10$^{6}$M$_{\odot}$)} & \colhead{(10$^{6}$M$_{\odot}$)} & \colhead{(10$^{6}$M$_{\odot}$)}
}
\startdata
NGC~3227 & 36.1$^{+21.0}_{-14.4}$ & 33.1$^{+76.3}_{-32.1}$ & 36$\pm$14 \\
NGC~3516 & 18.2$^{+8.1}_{-3.6}$   & 10.3$^{+7.7}_{-3.7}$   & 16.8$\pm$3.3 \\
NGC~4593 & 6.3$^{+5.6}_{-8.2}$    & 8.7$^{+21.3}_{-14.8}$  & 6.6$\pm$5.2 \\
\enddata
\end{deluxetable}


\clearpage

\begin{thebibliography}{}

\bibitem[Baribaud et al.(1994)]{bar94}  Baribaud, T., Salamanca, I., Alloin, D., \&
     Wagner, S. 1994, \aaps, 103, 121
\bibitem[Blandford \& McKee(1982)]{bla82}  Blandford, R. D., \& McKee, C. F. 1982,
     \apj, 255, 419
\bibitem[Chiang(2002)]{chi02}  Chiang, J. 2002, \apj, 572, 79
\bibitem[Czerny et al.(2001)]{cze01}  Czerny, B., Niko\l ajuk, M., Piasecki, M., \& 
     Kuraszkiewicz, J. 2001, \mnras, 325, 865
\bibitem[Di Nella et al.(1995)]{din95}  Di Nella, H., Garcia, A. M., Garnier, R., \&
     Paturel, G. 1995, \aaps, 113, 151
\bibitem[Dietrich et al.(1994)]{die94}  Dietrich, M., et al. 1994, \aap, 284, 33
\bibitem[Ferrarese(2002)]{fer02}  Ferrarese, L. 2002, in Current High-Energy Emission 
     around Black Holes, ed.\ C.-H.\ Lee (Singapore: World Scientific), in press 
     (astro-ph/0203047)
\bibitem[Ferrarese \& Merritt(2000)]{fer00}  Ferrarese, L., \& Merritt, D. 2000, \apj,
     539, L9
\bibitem[Ferrarese et al.(2001)]{fer01}  Ferrarese, L., Pogge, R. W., Peterson, B. M.,
     Merritt, D., Wandel, A., \& Joseph, C. L. 2001, \apj, 555, L79
\bibitem[Gebhardt et al.(2000a)]{geb00}  Gebhardt, K., et al. 2000a, \apj, 539, L13
\bibitem[Gebhardt et al.(2000b)]{geb00b}  Gebhardt, K., et al. 2000b, \apj, 543, L5
\bibitem[Ho(1999)]{ho99}  Ho, L. C. 1999, in Observational Evidence for Black Holes
     in the Universe, ed. S. K. Chakrabarti (Dordrecht: Kluwer), 157
\bibitem[Kaspi et al.(2000)]{kas00}  Kaspi, S., Smith, P. S., Netzer, H., Maoz, D.,
     Jannuzi, B. T., \& Giveon, U. 2000, \apj, 533, 631
\bibitem[Kazanas \& Nayakshin(2001)]{kaz01}  Kazanas, D., \& Nayakshin, S. 2001,
     \apj, 550, 655
\bibitem[Keel(1996)]{kee96}  Keel, W. C. 1996, \aj, 111, 696
\bibitem[Kollatschny \& Dietrich(1997)]{kol97}  Kollatschny, W., \& Dietrich, M.
     1997, \aap, 323, 5
\bibitem[Krolik(2001)]{kro01}  Krolik, J. H. 2001, \apj, 551, 72
\bibitem[Maoz \& Netzer(1989)]{mao89}  Maoz, D., \& Netzer, H. 1989, \mnras, 236, 21
\bibitem[Morales \& Fabian(2002)]{mor02}  Morales, R., \& Fabian, A. C. 2002, \mnras,
     329, 209
\bibitem[Nelson \& Whittle(1995)]{nel95}  Nelson, C. H., \& Whittle, M. 1995, \apjs,
     99, 67
\bibitem[Onken \& Peterson(2002)]{onk02}  Onken, C. A., \& Peterson, B. M. 2002,
     \apj, 572, 746
\bibitem[Peterson(1993)]{pet93}  Peterson, B. M. 1993, \pasp, 105, 247
\bibitem[Peterson \& Wandel(1999)]{pet99}  Peterson, B. M., \& Wandel, A. 1999,
     \apj, 521, L95
\bibitem[Peterson \& Wandel(2000)]{pet00}  Peterson, B. M., \& Wandel, A. 2000,
     \apj, 540, L13
\bibitem[Peterson et al.(1998)]{pet98}  Peterson, B. M., Wanders, I., Horne, K., 
     Collier, S., Alexander, T., Kaspi, S., \& Maoz, D. 1998, \pasp, 110, 660
\bibitem[Robinson(1994)]{rob94}  Robinson, A. 1994, in ASP Conf. Ser. 69, 
     Reverberation Mapping of the Broad-Line Region in Active Galactic Nuclei, ed.
     P. M. Gondhalekar, K. Horne, \& B. M. Peterson (San Francisco: ASP), 147
\bibitem[Rodr\'{\i}guez-Pascual et al.(1997)]{rod97}  Rodr\'{\i}guez-Pascual, P. M., 
     et al. 1997, \apjs, 110, 9
\bibitem[Salamanca et al.(1994)]{sal94}  Salamanca, I., et al. 1994, \aap, 282, 742
\bibitem[Santos-Lle\'{o} et al.(1995)]{san95}  Santos-Lle\'{o}, M., Clavel, J.,
     Barr, P., Glass, I. S., Pelat, D., Peterson, B. M., \& Reichert, G. 1995,
     \mnras, 274, 1
\bibitem[Schinnerer, Eckart, \& Tacconi(2000)]{sch00}  Schinnerer, E., Eckart, A.,
     \& Tacconi, L. J. 2000, \apj, 533, 826
\bibitem[Strauss et al.(1992)]{str92}  Strauss, M. A., Huchra, J. P., Davis, M., 
     Yahil, A., Fisher, K. B., \& Tonry, J. 1992, \apjs, 83, 29
\bibitem[Tremaine et al.(2002)]{tre02}  Tremaine, S., et al. 2002, \apj, 574, 740
\bibitem[van Groningen \& Wanders(1992)]{van92}  van Groningen, E., \& Wanders, I.
     1992, \pasp, 104, 700
\bibitem[WPM(1999)]{wan99}  Wandel, A., Peterson, B. M., \& Malkan, M. A. 1999,
     \apj, 526, 579 (WPM)
\bibitem[Wanders \& Horne(1994)]{wan94}  Wanders, I., \& Horne, K. 1994, \aap, 289, 76
\bibitem[Wanders et al.(1992)]{wan92}  Wanders, I., Peterson, B. M., Pogge, R. W., 
     DeRobertis, M. M., \& van Groningen, E. 1992, \aap, 266, 72
\bibitem[Wanders et al.(1993)]{wan93}  Wanders, I., et al. 1993, \aap, 269, 39
\bibitem[White \& Peterson(1994)]{whi94}  White, R. J., \& Peterson, B. M. 1994,
     \pasp, 106, 879
\bibitem[Winge et al.(1995)]{win95}  Winge, C., Peterson, B. M., Horne, K., Pogge, R. W.,
     Pastoriza, M. G., \& Storchi-Bergmann, T. 1995, \apj, 445, 680

\end{thebibliography}
\end{document}